\documentclass{aastex6} % AAS style

\slugcomment{} % remove "Draft version" text at top of page

\usepackage{graphicx}				% Use pdf, png, jpg, or eps§ with pdflatex; use eps in DVI mode
\usepackage{hyperref} 	% for URLs
\usepackage{url}
\usepackage{natbib}
\usepackage{amssymb}
\usepackage{booktabs}

\begin{document}
\title{The Astropy Problem} %

% The author entry is hacked a bit so that the affiliations can be listed at the bottom of the paper.

\author{
Demitri Muna\hyperlink{affil:1}{$^1$}$^{,}$\hyperlink{affil:2}{$^{2}$},
Michael Alexander,
Alice Allen\hyperlink{affil:3}{$^{3}$},
Richard Ashley\hyperlink{affil:4}{$^{4}$},
Daniel Asmus\hyperlink{affil:5}{$^{5}$},
Ruyman Azzollini\hyperlink{affil:6}{$^{6}$},
Michele Bannister\hyperlink{affil:7}{$^{7}$},
Rachael Beaton\hyperlink{affil:8}{$^{8}$},
Andrew Benson\hyperlink{affil:8}{$^{8}$},
G.~Bruce Berriman\hyperlink{affil:10}{$^{10}$},
Maciej Bilicki\hyperlink{affil:50}{$^{50}$},
Peter Boyce,
Joanna Bridge\hyperlink{affil:11}{$^{11}$},
Jan Cami\hyperlink{affil:12}{$^{12}$},
Eryn Cangi\hyperlink{affil:13}{$^{13}$},
Xian Chen\hyperlink{affil:14}{$^{14}$},
Nicholas Christiny,
Christopher Clark\hyperlink{affil:15}{$^{15}$},
Michelle Collins\hyperlink{affil:16}{$^{16}$},
Johan Comparat\hyperlink{affil:17}{$^{17}$},
Neil Cook\hyperlink{affil:18}{$^{18}$},
Darren Croton\hyperlink{affil:19}{$^{19}$},
Isak Delberth Davids\hyperlink{affil:20}{$^{20}$},
\'{E}ric Depagne\hyperlink{affil:21}{$^{21}$},
John Donor\hyperlink{affil:22}{$^{22}$},
Leonardo A. dos Santos\hyperlink{affil:23}{$^{23}$},
Stephanie Douglas\hyperlink{affil:24}{$^{24}$},
Alan Du\hyperlink{affil:24}{$^{24}$},
Meredith Durbin\hyperlink{affil:25}{$^{25}$},
Dawn Erb\hyperlink{affil:26}{$^{26}$},
Daniel Faes\hyperlink{affil:27}{$^{27}$},
J.~G. Fern\'{a}ndez-Trincado\hyperlink{affil:28}{$^{28}$},
Anthony Foley,
Sotiria Fotopoulou\hyperlink{affil:29}{$^{29}$},
S{\o}ren Frimann\hyperlink{affil:30}{$^{30}$},
Peter Frinchaboy\hyperlink{affil:22}{$^{22}$},
Rafael Garcia-Dias\hyperlink{affil:49}{$^{49}$},
Artur Gawryszczak\hyperlink{affil:103}{$^{103}$},
Elizabeth George\hyperlink{affil:32}{$^{32}$},
Sebastian Gonzalez\hyperlink{affil:33}{$^{33}$},
Karl Gordon\hyperlink{affil:34}{$^{34}$},
Nicholas Gorgone\hyperlink{affil:35}{$^{35}$},
Catherine Gosmeyer\hyperlink{affil:34}{$^{34}$},
Katie Grasha\hyperlink{affil:36}{$^{36}$},
Perry Greenfield\hyperlink{affil:34}{$^{34}$},
Rebekka Grellmann\hyperlink{affil:37}{$^{37}$},
James Guillochon,
Mark Gurwell\hyperlink{affil:38}{$^{38}$},
Marcel Haas,
Alex Hagen\hyperlink{affil:11}{$^{11}$},
Daryl Haggard\hyperlink{affil:39}{$^{39}$},
Tim Haines\hyperlink{affil:40}{$^{40}$},
Patrick Hall\hyperlink{affil:41}{$^{41}$},
Wojciech Hellwing\hyperlink{affil:42}{$^{42}$},
Edmund Christian Herenz\hyperlink{affil:99}{$^{99}$},
Samuel Hinton\hyperlink{affil:43}{$^{43}$},
Renee Hlozek\hyperlink{affil:101}{$^{101}$},
John Hoffman\hyperlink{affil:44}{$^{44}$},
Derek Holman,
Benne Willem Holwerda\hyperlink{affil:45}{$^{45}$}$^{,}$\hyperlink{affil:96}{$^{96}$},
Anthony Horton,
Cameron Hummels\hyperlink{affil:46}{$^{46}$},
Daniel Jacobs\hyperlink{affil:47}{$^{47}$},
Jens Juel Jensen\hyperlink{affil:48}{$^{48}$},
David Jones\hyperlink{affil:49}{$^{49}$},
Arna Karick,
Luke Kelley\hyperlink{affil:38}{$^{38}$},
Matthew Kenworthy\hyperlink{affil:50}{$^{50}$},
Ben Kitchener,
Dominik Klaes\hyperlink{affil:51}{$^{51}$},
Saul Kohn\hyperlink{affil:52}{$^{52}$},
Piotr Konorski\hyperlink{affil:100}{$^{100}$},
Coleman Krawczyk\hyperlink{affil:42}{$^{42}$},
Kyler Kuehn\hyperlink{affil:53}{$^{53}$},
Teet Kuutma\hyperlink{affil:54}{$^{54}$},
Michael T. Lam\hyperlink{affil:55}{$^{55}$},
Richard Lane\hyperlink{affil:14}{$^{14}$},
Jochen Liske\hyperlink{affil:102}{$^{102}$},
Diego Lopez-Camara\hyperlink{affil:56}{$^{56}$},
Katherine Mack\hyperlink{affil:57}{$^{57}$},
Sam Mangham\hyperlink{affil:58}{$^{58}$},
Qingqing Mao,
David J.~E. Marsh\hyperlink{affil:59}{$^{59}$},
Cecilia Mateu\hyperlink{affil:60}{$^{60}$},
Lo\"{i}c Maurin\hyperlink{affil:14}{$^{14}$},
James McCormac\hyperlink{affil:4}{$^{4}$},
Ivelina Momcheva\hyperlink{affil:34}{$^{34}$},
Hektor Monteiro\hyperlink{affil:61}{$^{61}$},
Michael Mueller\hyperlink{affil:62}{$^{62}$},
Roberto Munoz\hyperlink{affil:14}{$^{14}$},
Rohan Naidu\hyperlink{affil:63}{$^{63}$},
Nicholas Nelson\hyperlink{affil:64}{$^{64}$},
Christian Nitschelm\hyperlink{affil:65}{$^{65}$},
Chris North \hyperlink{affil:15}{$^{15}$},
Juan Nunez-Iglesias\hyperlink{affil:66}{$^{66}$},
Sara Ogaz\hyperlink{affil:34}{$^{34}$},
Russell Owen\hyperlink{affil:25}{$^{25}$},
John Parejko\hyperlink{affil:25}{$^{25}$},
Vera Patr'cio\hyperlink{affil:67}{$^{67}$},
Joshua Pepper\hyperlink{affil:68}{$^{68}$},
Marshall Perrin\hyperlink{affil:34}{$^{34}$},
Timothy Pickering\hyperlink{affil:34}{$^{34}$},
Jennifer Piscionere\hyperlink{affil:19}{$^{19}$},
Richard Pogge\hyperlink{affil:1}{$^{1}$}$^{,}$\hyperlink{affil:2}{$^{2}$},
Radek Poleski\hyperlink{affil:2}{$^{2}$},
Alkistis Pourtsidou\hyperlink{affil:42}{$^{42}$},
Adrian M. Price-Whelan\hyperlink{affil:69}{$^{69}$},
Meredith L. Rawls\hyperlink{affil:25}{$^{25}$},
Shaun Read\hyperlink{affil:18}{$^{18}$},
Glen Rees\hyperlink{affil:70}{$^{70}$},
Hanno Rein\hyperlink{affil:71}{$^{71}$},
Thomas Rice\hyperlink{affil:72}{$^{72}$},
Signe Riemer-S\o{}rensen\hyperlink{affil:73}{$^{73}$},
Naum Rusomarov\hyperlink{affil:74}{$^{74}$},
Sebastian F. Sanchez\hyperlink{affil:56}{$^{56}$},
Miguel Santander-Garc\'{i}a\hyperlink{affil:75}{$^{75}$},
Gal Sarid\hyperlink{affil:76}{$^{76}$},
William Schoenell\hyperlink{affil:27}{$^{27}$},
Aleks Scholz\hyperlink{affil:77}{$^{77}$},
Robert L. Schuhmann\hyperlink{affil:78}{$^{78}$},
William Schuster\hyperlink{affil:79}{$^{79}$},
Peter Scicluna,
Marja Seidel\hyperlink{affil:8}{$^{8}$},
Lijing Shao\hyperlink{affil:80}{$^{80}$},
Pranav Sharma\hyperlink{affil:81}{$^{81}$},
Aleksandar Shulevski\hyperlink{affil:82}{$^{82}$},
David Shupe\hyperlink{affil:10}{$^{10}$},
Crist\'{o}bal Sif\'{o}n\hyperlink{affil:44}{$^{44}$},
Brooke Simmons\hyperlink{affil:83}{$^{83}$},
Manodeep Sinha\hyperlink{affil:19}{$^{19}$},
Ian Skillen\hyperlink{affil:104}{$^{104}$},
Bjoern Soergel\hyperlink{affil:84}{$^{84}$},
Thomas Spriggs\hyperlink{affil:18}{$^{18}$},
Sundar Srinivasan\hyperlink{affil:85}{$^{85}$},
Abigail Stevens\hyperlink{affil:86}{$^{86}$},
Ole Streicher\hyperlink{affil:87}{$^{87}$},
Eric Suchyta\hyperlink{affil:88}{$^{88}$},
Joshua Tan\hyperlink{affil:14}{$^{14}$},
O.~Grace Telford\hyperlink{affil:25}{$^{25}$},
Romain Thomas\hyperlink{affil:9}{$^{9}$},
Chiara Tonini\hyperlink{affil:89}{$^{89}$},
Grant Tremblay\hyperlink{affil:98}{$^{98}$},
Sarah Tuttle\hyperlink{affil:25}{$^{25}$},
Tanya Urrutia\hyperlink{affil:87}{$^{87}$},
Sam Vaughan\hyperlink{affil:31}{$^{31}$},
Miguel  Verdugo\hyperlink{affil:97}{$^{97}$},
Alexander Wagner\hyperlink{affil:90}{$^{90}$},
Josh Walawender\hyperlink{affil:91}{$^{91}$},
Andrew Wetzel\hyperlink{affil:8}{$^{8}$}$^{,}$\hyperlink{affil:92}{$^{92}$}$^{,}$\hyperlink{affil:93}{$^{93}$},
Kyle Willett,
Peter K.~G. Williams\hyperlink{affil:38}{$^{38}$},
Guang Yang\hyperlink{affil:11}{$^{11}$},
Guangtun Zhu\hyperlink{affil:94}{$^{94}$},
Andrea Zonca\hyperlink{affil:95}{$^{95}$}\\
\bigskip
Affiliations can be found at the end of the paper.\\
\smallskip
} % end author

\date{10 October 2016}

\begin{abstract}The Astropy Project (\url{http://astropy.org}) is, in its own words, ``a community effort to develop a single core package for Astronomy in Python and foster interoperability between Python astronomy packages." For five years this project has been managed, written, and operated as a grassroots, self-organized, almost entirely volunteer effort while the software is used by the majority of the astronomical community. Despite this, the project has always been and remains to this day effectively unfunded. Further, contributors receive little or no formal recognition for creating and supporting what is now critical software. This paper explores the problem in detail, outlines possible solutions to correct this, and presents a few suggestions on how to address the sustainability of general purpose astronomical software.

\bigskip

\textit{N.B.: This is not an official paper of nor officially endorsed by the Astropy Project, rather it is a reflection of the opinions of the authors.}

\end{abstract}

\section{Astronomy and Python}

\subsection{Python, Inexorably}
Major shifts in a community are not always easy to identify years before they gain significant momentum, but in some cases they can be quite evident. As an arbitrary date, let's go back six years to 2010. If you had asked most astronomers then if the community was going to move to Python, the answer would have been a very clear ``yes". At the time very few astronomers actually knew Python. Many were hesitant to invest the time required to learn a new language, though many were interested in learning. There were few astronomical Python libraries available at the time, but the canaries in the astronomy software coal mine were already developing in it.

One major (but not exclusive) driver of the need for change, and probably the most significant, was widespread dissatisfaction with IDL. Those who strongly feel that scientific software and analysis should be open are opposed to the high license fees (or really, any cost) required to run the code. Astronomers are not the primary users of the language; certainly not the community that guides or drives its development. Most additions to the platform over the past fifteen years have focussed on the components used to build graphical user interfaces, an aspect that our community barely uses. While IDL provided obvious utility in the 1990s in scripting and plotting over C and FORTRAN, these advantages evaporated with the advent of modern scripting and visualization languages. Python, coupled with the rapidly maturing NumPy numerical library \citep{numpy}, was the unambiguous, natural successor.

There were two barriers to adopting Python: 1) astronomers would be required to learn a new language, and 2) the substantial IDL libraries that astronomers rely on daily needed to be replaced with Python versions. To help address the first problem, the lead author started the SciCoder workshop in 2010,\footnote{OK, 2010 was a semi-arbitrary date.} five days dedicated to teaching early career astronomers Python and modern software development techniques, offered annually and at AAS workshops. Demand for this well exceeded supply. At the early workshops, virtually none of the attendees had ever used Python; most used IDL with some C or FORTRAN. Software Carpentry\footnote{\url{http://software-carpentry.org}} is another example of an effort to address this need. Of course, many people decided to learn Python on their own.

By this time, however, funding for general purpose software development had all but disappeared. Funding for the development of packages most used by astronomers such as NOAO's IRAF \citep{iraf}, the Smithsonian's SAOImage DS9 \citep{ds9}, the (former) PPARC's Starlink Project (\cite{starlink}, funding cut to zero in 2005), and NRAO's AIPS \citep{aips} was either completely cut or has trickled to the bare minimum for maintenance purposes. None of the institutions that had funded and developed the everyday tools that astronomers rely on were driving new software development, with the notable exception of Space Telescope Science Institute (STScI) (see below). We are specifically referring to general purpose tools; naturally survey or mission-specific software pipelines continue to be written. It would be one thing if there were widespread disagreement on the next direction of software development, but given that the path forward was clear and almost universally agreed upon, it would have been reasonable to expect that some institution -- one that had led in software development previously or somewhere new -- would have taken on the task of writing the next generation of tools, in Python, that astronomers would be needing. It is not unreasonable to say that no new, major, general-purpose software packages have been created since that time, while we continue to increasingly struggle to use and maintain tools effectively written fifteen or twenty years ago (and some even older).

\subsection{Enter Astropy}
Actually, that last statement is not true; there are two new, notable software packages that have appeared. One is yt \citep{2011ApJS..192....9T}, started by Matt Turk and developed by a small community of astronomers. yt is an impressive, modern package for analyzing volumetric, multi-resolution data, primarily used by astronomers working with data from large simulations. While somewhat specialized, this package certainly counts. The other major software package is of course Astropy.

The foundations for Astropy began in 1998 at the Space Telescope Science Institute. Difficulties in working with IRAF (both technical and managerial) motivated a move to Python, first through PyRAF \citep{pyraf} and then through early development of tools that were missing from the language. These included support for numerical arrays (numarray \citep{numarray} which directly led to NumPy), plotting (matplotlib, developed with the original author John Hunter \citep{Hunter:2007}), and FITS handling via PyFITS \citep{pyfits}. All of this development was in direct support of STScI missions (e.g.~Hubble Space Telescope, James Webb Space Telescope); it had to be continually argued that the software be written to be generally applicable to other telescopes.

% -- IRAF increasingly difficult to maintain - people don't realize this
% -- 
Eight years ago, a handful of astronomers who no longer had patience with how outdated existing tools were and wanted to move to a modern environment began to do so. The first barrier was the lack of astronomy libraries in Python, so we began to write them. This effort was not organized; many individuals began writing the packages they needed largely independently and in their spare time. We will highlight two in particular. Erik Tollerud created Astropysics \citep{astropysics} in 2008, an impressive package that implements coordinate transformations, cosmological calculations, image processing tools, data visualization, and much more. At the time he was a graduate student at UC Irvine. At one point he did a survey of astronomical software available on personal home pages of various people and counted fourteen independent Python packages to perform coordinate system transformations. While much of the functionality overlapped, each scratched a particular itch that the others did not.\footnote{The primary author will confess to have begun work with Adrian Price-Whelan on what would have become the fifteenth before becoming aware of all of the others.} Part of the problem was one of discovery; unless one spent time actively looking for these packages, it was very easy to be unaware of them. (This was before GitHub became the standard for open source code distribution.) Thomas Robitaille, at the time a Spitzer Postdoctoral Fellow at the Harvard-Smithsonian Center for Astrophysics, also began writing astronomical software in 2008. He, along with Eli Bressert, wrote APLpy \citep{aplpy}, a plotting library for astronomical data.

The Astropy project was formed in 2011 to create a single, definitive Python package for astronomy. Much of the software developed by STScI highlighted above was incorporated into the new development, updated to be consistent as a single, coherent code base. While this was a collaborative grassroots effort by roughly one to two dozen core developers, we highlight Tollerud and Robitaille specifically as they played a strong --- and critical --- role in organizing, managing, and mediating the community. It was recognized that developing many independent packages was unsustainable and wasted a tremendous amount of work from highly skilled astronomer-programmers. The following aims were recognized from the beginning:

\begin{itemize}
\item{the library should be modular rather than monolithic}
\item{code should be built on top of a set of common base classes}
\item{the application program interface (API) should be clean, clear, and easy to use}
\item{code should be fully object-oriented instead of a straight translation of existing libraries}
\item{code should work and integrate with the existing Python ecosystem}
\end{itemize}

\noindent
The first commit to Astropy on GitHub occurred on 25 July 2011, and the first in-person organizational/steering/coding sprint meeting occurred on 12-14 October 2011. Since then, there have been 191 distinct contributors to the project. As of July 2016, it contains over 215,000 lines of code. (More statistics are listed in Table \ref{table:astropy-stats}.) Highlights of the library include handling coordinate systems, times and dates, modeling, FITS, ASCII, and VO file access, cosmological calculations, data visualization, and much more. The details of the beginnings of Astropy are interesting, but beyond the scope of this paper.\footnote{The interested reader is directed to \url{https://mail.scipy.org/pipermail/astropy/2011-June/thread.html\#1380}.}

\section{The Astropy Problem}

Astropy is an unqualified success for the astronomical community. It funnels and has guided consistent, enormous efforts of dozens of astronomers from across the world to produce a first class, high quality, user friendly, fully documented software library and toolset for more than five years. What is its funding model? This is easy to describe: it doesn't have one. Beyond a few developers at STScI (who have since left astronomy), almost none have been paid at all for their efforts. There are no Astropy positions in any academic or scientific institution. Much of the development has been undertaken overwhelmingly by graduate students and postdocs, with some representation from undergraduate students and faculty members, in their spare time.

Whose responsibility is it to pay for this development? The following is based on discussions with many people over the past few years. Employees of NASA institutions say that each dollar spent must be allocated to a specific project or mission; money cannot be spent on general purpose, community efforts. Those at academic institutions say their primary mission is education, not software development. Other scientific institutions (some of which have funded general purpose software in the past) have noted that their responsibility is to operate and run the telescopes and data archives under their remit. Individual surveys received the money they did (either through government or private funding) to deliver the science products promised, not to develop community software. We would like to specifically highlight feedback the first author has heard second-hand from two different senior NSF reviewers. The first is that ``the NSF hasn't funded software development for many years now, yet software continues to be written, so...?" The second is that the NSF ``would consider funding software, but only once it has reached the point where it's very mature and has a large number of users." These opinions reflect a gross misunderstanding of the amount of time, effort, and expertise it takes to develop software, let alone specialized, scientific software. Further, it values these efforts (and frankly, the people performing them) at exactly zero. Entire companies filled with people working full time\footnote{None of which are trying to pursue active research and publications at the same time.} are formed to create software products that are mature and have a large number of users. To expect this of someone in an academic environment with no guarantee of financial or career reward is ridiculous on the face of it.

Individually, the positions of the NASA managers, the academics, the funding bodies, the surveys, and the managers of national laboratories seem reasonable and, within their scope and politics, justifiable. (Except for the two specific opinions listed above.) However, together these bodies represent the whole of the astronomical community. Put explicitly: represented by those that have the ability, the entire astronomical community has decided that Astropy, and general purpose software development, is not something that it is willing to fund. Why should they be obligated to? 

The answer is that every one of these bodies not only uses Astropy, but critically depends on it. All NASA mission pipelines, from the venerable Hubble Space Telescope to the upcoming WFIRST, use Astropy, and increasingly so every day. Nearly all national laboratories, surveys, and telescopes depend on Astropy. If academic students and staff are not explicitly typing ``import astropy" as the first line of their analysis codes, they are using data collected, reduced, or analyzed by code that has. Nearly every single astronomer today uses or benefits directly from Astropy.  The position of \textit{any} who claims they are not the one to support Astropy is in fact \textit{not} justifiable. The larger the organization, the less justifiable it is.

How much money has Astropy saved these institutions? To estimate this, we used a program called SLOCCount\footnote{\url{http://www.dwheeler.com/sloccount/}} (Source Lines of Code), a common, industry standard tool that estimates the cost in time and money to develop a project based on the source code. The default software cost estimation model, the Constructive Cost Model (COCOMO, \cite{1981see..book.....B}) was used to avoid introducing any biases. An average salary of \$100,000 with a 1.5x multiplier for overheads was used. This is a rough number; junior astronomers would make less, more senior astronomers would make more, and overheads vary as well. The value is roughly comparable to salaries of highly skilled software developers in industry, which is reasonable given that most of the Astropy developers either have or would soon have postgraduate degrees. The fact that our community does not always compensate at these levels does not diminish the true value of the contribution. (The reader is of course free to either run the cost estimation software with their own numbers or scale based on how they value the effort.)

% Requires the booktabs if the memoir class is not being used
\begin{table}[t]
   \centering
   %\topcaption{Table captions are better up top} % requires the topcapt package
   \begin{tabular}{@{} lr @{}} % Column formatting, @{} suppresses leading/trailing space
      \toprule
      \multicolumn{2}{c}{Astropy Codebase Statistics} \\
      \midrule
        First commit date & 25 July 2011 \\
        Number of distinct contributors &	191 \\
        Number of commits & 15,475 \\
        Lines of Python code & 200,295 \\
        Lines of Python code (core package) & 125,737 \\
        Lines of C code & 13,957 \\
        Lines of shell code & 1,065 \\
        Total lines of code & 215,317 \\
        Development effort estimate (person-years) & 56.33 \\
        Schedule estimate & 2.48 years with 22.72 developers \\
        \midrule
        Estimated cost & \$8,449,937\\
      \bottomrule
   \end{tabular}
   \caption{Statistics are based on the Astropy repository as of July 2016, version v1.2. All repositories under the Astropy GitHub account have been included. External C libraries (cfitsio, ERFA, expat, and wcslib) have been excluded. Cost and development estimates generated using David A. Wheeler's ``SLOCCount" software. }
   \label{table:astropy-stats}
\end{table}

The estimated cost to develop Astropy is \$8.5M. The model estimates that it would take nearly 23 full-time developers two and a half years to reproduce what currently exists in the repository. Given that the project is five years old, the cost to develop Astropy is \$1.7M/year. Or rather, the uncompensated effort donated to the community by the 191 distinct contributors is worth \$1.7M/year. (The time period covers the lifetime of Astropy, not the start date the code included began to be written.) This value is conservative: it doesn't include talks, tutorials, the extensive and detailed documentation, and technical support that is also provided. Astropy is a huge success for the aforementioned institutions -- they didn't have to pay for any of it (again, with the exception of a few FTEs from STScI which are no longer present). But the reward for contributors is just one paper,\footnote{And we all know how much a purely software paper is valued in our community.} roughly 400 citations directly related to the work, little recognition, and precious few career-track software development positions in the field. Additionally, there is an expectation that Astropy will continue to be developed, have new features added and bugs fixed, and that it will be integrated into more and more mission and research pipeline, analysis, and visualization code. For much of the community, software is something that just ``happens" and is expected to be free. However, there is a cost and it is clear who is paying it. The early career astronomers who contribute the lion's share of the effort do so at the expense of their research and publication output. The time they sacrifice to work on Astropy is not considered by hiring committees and does not improve their prospects on the academic job market; in fact, it's detrimental. Of course, many use code they write to help with their own research, but even then it's not their job to provide the community with software that has long term viability; this additional effort is well beyond one's specific research needs.

It's worth specifying what the Astropy community provides, particularly for those who are only familiar with the end product. Writing software for a few thousand users is vastly different than writing an analysis script for oneself or a few people. The software must be cleanly designed, provide clear error messages, and come with example code. It must be well commented so that others may easily fix or enhance existing code. It must be robust enough to run in as many environments as possible. It must run on three different platforms (macOS, Linux, Windows) over several different operating systems on each. It must be trustworthy, which requires a significant amount of user testing and code-based unit test suites (code contributed to Astropy core is reviewed by at least one other developer before it is accepted). It must be performant, which requires code optimization and an understanding of the detailed minutiae of both Python and the underlying operating systems. It requires extensive documentation, a significant effort by itself. Bugs must be addressed quickly. Users of Astropy enjoy an unprecedented level of support; anyone can request help or advice from the Astropy user mailing list\footnote{https://mail.scipy.org/mailman/listinfo/astropy} and receive a response--- often within minutes. (Even barring all other development, consider for a moment what this level of support would cost the STScI or NOAO to provide.) Not one but two Astropy releases are actively maintained -- one for new features, and one that provides long term support where bugs are fixed but not new functionality. The time required for maintenance scales with the number of lines of code and competes with new development. New feature requests are discussed with users and developed whenever possible. Code patches submitted are tested, modified to be consistent with the exiting base and coding style (when necessary), and documentation updated. Beyond the core Astropy package, the community develops, supports, and manages numerous affiliated packages. On top of this the project is managed as professionally as one would expect from any software company, integrating work from nearly two hundred contributors, while regularly producing new releases and functionality used by the entire astronomical community. All while being almost entirely unfunded.

\section{Missed Opportunities}
One of the big successes of the Astropy organizers is that they were able to take a wide range of astronomers -- seasoned developers, people who were new to writing software, people just learning Python, people who didn't feel comfortable coding but wanted to contribute -- and bring them together to create a single library. They did this for people who were scattered across the globe while they themselves worked at different institutions, continents, and time zones. The developers were quick and early adopters of many new technologies, including embracing the community development platform of GitHub, adopting continuous integration, unit testing, code review, and object-oriented code, while performing all this in a completely open and transparent manner. The needs of the \textit{users} were placed higher than the needs of any particular project, survey, or mission.

The reality is that no current astronomy institution could have created the equivalent to Astropy, even if funding were available (despite its early origins at STScI which undoubtedly laid a foundation for the grass roots work, the success and sustained momentum of the project is attributable to the developers, organizers, and the community created around them). This is partly due to institutional culture, partly due to management, and partly due to the extreme competition for funds. If reasonable funding were made available, likely it would have taken the form of competing efforts of more than one institution with significant overlap in functionality. By being independent, Astropy development is more agile than most other efforts in the astronomy community, but is slower than it could be with a sustained commitment of financial support. It would be easy to say in 2011 that it was not clear that Astropy would have been successful, and that it would have been risky to provide significant funding to graduate students for an unproven project. It might be arguable at which point the project evolved to one that became indispensable and critical to the community, but it is indisputable that that line has long since been crossed.

What should have happened is a clear recognition by the community that something significant has been accomplished, something outside of the traditional structure. The response should have been: ``Tollerud and Robitaille et al., here is a big pile of money to support, continue, and expand what you are doing. We won't tell you how to do it or attach strings; you clearly know what you are doing." A compelling argument was made to me in this regard. We measure success in science by how many people use the results of or are impacted by our work. We are encouraged to publish our results so that others build upon them, which are measured by the number of citations. While there is an Astropy paper, there is no doubt that the library is used by far more people than actually cite it -- probably close to an order of magnitude. The number of astronomers whose work was built with the assistance of Astropy rivals or exceeds the most well-cited papers. This alone should have led to the equivalent of tenured positions for the principal developers of Astropy -- certainly tenured positions have been awarded to those less impactful. Such positions in software development don't exist in our community.

The ideas of ``data science" and ``big data" began to gain significant traction at the same time Astropy began in earnest as a project. Notable amounts of money became available to pursue these interests. Astropy and similar efforts would have been the natural kinds of projects to support. These new data science initiatives were perfectly timed to embrace, support, and extend community development without existing constraints (funding, management, or otherwise); they would have been a natural home to projects like Astropy and the establishment of new, similar projects. While there are examples to the contrary, in many instances this new influx of funds have been used to simply support existing work and positions. As more institutions are coming into this sector, they have a unique, but as yet unrealized opportunity to take a leadership role in the community.

As an example, development on the LSST data pipeline began as an open source project in 2004 \citep{2004AAS...20510811A} written  in a mix of C++ and Python, nine years before Astropy development began. Since that time, the community has overwhelmingly chosen to use Astropy as the next generation of astronomical software over LSST's library. There are several reasons for this (including political), but two primary ones are community involvement and high quality APIs and documentation.

Astropy grew directly from community need and with very close interaction with the community, while LSST's software was guided by the specific needs of a single survey. This is an extremely important distinction -- placing the users above the project -- that is nearly unique to Astropy in astronomy. The two are not strictly mutually exclusive: software can be developed for a project while directly benefitting the community at large. SAOImage DS9 is an example of this; it was specifically funded and written to be the visualization tool for the NASA Chandra X-ray observatory (hence the software's connection to the Harvard-Smithsonian Center for Astrophysics), but clearly its use has far exceeded this purpose.

Astropy also provides an excellent, easy to use interface in the form of good APIs, clear and thorough documentation, and varied usage examples. This allows new users to quickly get up to speed on how to use the software, while preventing developer friction because of unclear or unpleasant interfaces. Good APIs attract good developers and keep them happy when they work on code. Happy developers are more productive and produce better code.

The LSST management has recently recognized the value of integrating Astropy into their software and held a coordination meeting with several core Astropy developers. This is an excellent first step and an acknowledgement of not only how much the community has embraced Astropy, but the importance of high quality, easy to use APIs and documentation. Today, it would be foolish to develop significant astronomy software without incorporating or linking to Astropy in some fashion. LSST identified a number of areas where the use of Astropy could benefit LSST's software and where LSST could contribute code to Astropy, and made several requests and proposals for modifications to Astropy software to integrate more easily with LSST's software \citep{LSST&Astropy}. There was no offer of funding for this additional work.

\section{Moving Forward}
It is imperative that Astropy be properly and sustainably funded. It is imperative that general purpose software tools and libraries for astronomers be funded. The current code infrastructure (e.g.~IRAF, DS9) is rapidly crumbling, was never designed or intended to have the longevity being asked of it, and is generations behind modern design and techniques. Our tools are insufficient to cope with the flood of data we have today, let alone what we will have a few years from now. There is no institution within the astronomy community who does not have some responsibility to support these efforts as all benefit significantly from them and have for years. While there are policy or political roadblocks, it is the responsibility of the appropriate managers or directors to address them directly and remove them, not defer to them. Funding must be made available not only to support today's software, but software that will be needed several years from now. Career positions that offer stability and respectable salaries must be created. Below we propose some possible solutions, but more solutions and efforts will be needed than are listed here.

\subsection{The Astropy Subscription}
We introduce here the concept of an Astropy subscription. As a community we are used to paying for IDL licenses. These costs are budgeted every year and included in grant proposals. We are actually quite familiar with paying for software subscriptions, so this will fit within existing infrastructure at every level. University departments would pay a subscription fee based on the size of the department. Larger institutions would consequently pay higher fees, perhaps based on number of users, budget size, or some other equitable metric. The subscription fees would be paid purely on a volunteer basis; there would be no license server, no restriction of the use of the code in any way, or literally any modification of how Astropy is currently distributed.  The subscription fee should be given as a way to provide direct support for the software and continued maintenance that each institution, survey, and mission is benefitting from. Contributors would be recognized on an annual basis on an Astropy web page. The annual total should be designed to approximately match the cost of development; from above, \$2M/year is a good starting point.

There is precedent for the ``professional moral obligation" argument. In the United States, lawyers (both individuals and law firms) are expected by the American Bar Association to contribute a minimum of fifty hours of legal services on a purely pro bono basis.\footnote{\url{http://www.americanbar.org/groups/probono_public_service/policy/aba_model_rule_6_1.html}} It's not directly enforceable -- one would not lose their license to practice for not doing so -- but it is a community expectation quantitatively set by a professional organization. It's reasonable to imagine that lawyers who do not volunteer this effort might be looked upon unfavorably by the community. We propose here that bodies such as the NSF, AAS, or the Decadal Plan committees work with the community to recommend the level of contribution to fund general software development projects. 

\subsection{Full Time Developers}
There are many larger scale surveys, missions, and projects within the community. As detailed above, the larger the entity, the greater benefit they derive from Astropy. We propose setting a financial water mark level above which will trigger a contribution to Astropy. For the sake of argument here, we choose \$40M (no special weight or significance should be attached to this number; it is selected as it is the approximate cost of the Sloan Digital Sky Survey). Once a project reaches this level, they will be required to hire a full time Astropy developer. \$150,000 is a small amount compared to the full budget. Naturally, the project should directly benefit from this: the developer would work within and be hosted by the survey/mission, and be a liaison between it and Astropy. They could work on Astropy code that would directly benefit the survey, an affiliated package, a custom visualization, etc., as long as the code was written for Astropy and could be used or easily adapted for use by the general community. The survey would not be allowed to pull this person to work on a mission-specific pipeline, for example. Of course, some percentage of their time would be given to the scientist to work on their own research, which would (should) naturally align with the survey. Two developers could be hired at twice the water mark, and so on. The exact scale is certainly open for debate, but the main point is that larger surveys should directly hire developers who would then work on Astropy or other community software. In fact, this model is very similar to that of the Starlink Project, where one full time Starlink developer was hosted at each astronomy department in the UK. This was a highly successful model which produced a significant amount of software used by nearly all of the Commonwealth astronomers. The project lost all funding not due to the lack of success, but pressure from larger departments who wanted to divert the development efforts to their specific needs rather than those of the wider community.

\subsection{The Decadal Survey}
The time for the Decadal Survey to address the problems outlined in this paper is now. These problems have been raised and discussed at all levels for years now (e.g.~\cite{2009astro2010P..61W}), but are only acknowledged instead of solutions being provided. We propose that the Decadal Survey include language to codify the Astropy subscription fee, allocate funds for the development of community software tools, and create the general software developer positions within astronomy surveys and missions as described above. Developers (i.e.~not managers) of astronomical software and computational technology more than deserve a seat at the table. We strongly suggest that all Decadal Survey meetings include representatives from both Astropy specifically and astronomical software development in general. The problem of software development was discussed in the context of the last Decadal Survey. Quoting James Turner from one of the earliest mailing list threads\footnote{\url{https://mail.scipy.org/pipermail/astropy/2011-June/001400.html}} that led to the creation of Astropy, dated 13 June 2011:

\begin{quote}
A couple of years ago, a number of us at the observatories submitted a
white paper to the Decadal Survey, pointing out the need for more
co-ordinated funding so that we can have people who focus on cross-institutional
platform development \& support. The report from the
committee did give a nod to our concerns and their importance, but
stopped short of making any recommendation, which basically means ``good luck with that."
\end{quote}

\noindent
The authors of the last Decadal Survey committed the community to not funding general utility astronomical software for a decade. Let's do better this time.

\section{Conclusion}

The aim of this paper is to point out a serious problem that exists in the astronomical community and give it a name. It is by no means limited to Astropy, but extends to community software that has been neglected and software that is needed and remains unwritten. It is not a problem that people are unaware of; it has been discussed at every level for many years now, but nearly all in the community who have had the resources to address it have either chosen not to do so or do not feel it is their responsibility. As the entire community benefits -- significantly and critically -- from the efforts of several dozen developers, it is in fact everyone's responsibility. This responsibility is not limited to US-based institutions and funding bodies: astronomy is a global community, and this support should be come from and be incorporated into the funding models around the world. The problem will not solve itself. It requires action, now. The community receives (and now expects) enormous utility from a number of developers without support, compensation, or career paths. The status quo is untenable and unethical. It is also short sighted; we currently lack the tools needed to fully analyze the sheer volume of data that is publicly available today. General purpose, community software has not been sufficiently funded for well over a decade -- the productivity lost in that time is significant.

The developers of Astropy and other software that push the boundaries of data science in astronomy are highly skilled members of our own community. They are most often the same ones who write and maintain the software infrastructure that run our surveys and space missions. Overwhelmingly they have taken it upon themselves to acquire the non-trivial skills needed to create these tools. It is only reasonable that the community in turn not only reward these efforts, but respect, encourage, and enable them to innovate further. 

\bigskip

\section*{Citing Astropy}
Every user can and should help to support the Astropy project. If you use Astropy in any part of a paper you write -- if any of your code includes the line ``import astropy" -- please be sure to cite the paper \citep{2013A&A...558A..33A}.  We know there are many, many more than 400 of you that use this!

\bigskip

\section*{About This Paper}
This paper is not an official paper of nor officially endorsed by the Astropy Project. It is a reflection of the opinion of the authors. Some core organizers of Astropy were specifically not invited to appear as co-authors to avoid creating a conflict of interest; however, their feedback and opinions were solicited and incorporated into the paper, and there were no objections to its publication. The paper was written by Demitri Muna who invited members of the astronomical community to be added as co-signers as a show of support.

\bibliography{astropy_problem}

\begin{thebibliography}{}
\expandafter\ifx\csname natexlab\endcsname\relax\def\natexlab#1{#1}\fi

\bibitem[{{Astropy Collaboration} {et~al.}(2013){Astropy Collaboration},
  {Robitaille}, {Tollerud}, {Greenfield}, {Droettboom}, {Bray}, {Aldcroft},
  {Davis}, {Ginsburg}, {Price-Whelan}, {Kerzendorf}, {Conley}, {Crighton},
  {Barbary}, {Muna}, {Ferguson}, {Grollier}, {Parikh}, {Nair}, {Unther},
  {Deil}, {Woillez}, {Conseil}, {Kramer}, {Turner}, {Singer}, {Fox}, {Weaver},
  {Zabalza}, {Edwards}, {Azalee Bostroem}, {Burke}, {Casey}, {Crawford},
  {Dencheva}, {Ely}, {Jenness}, {Labrie}, {Lim}, {Pierfederici}, {Pontzen},
  {Ptak}, {Refsdal}, {Servillat}, \& {Streicher}}]{2013A&A...558A..33A}
{Astropy Collaboration}, {Robitaille}, T.~P., {Tollerud}, E.~J., {et~al.} 2013,
  {Astronomy \& Astrophysics}, 558, A33

\bibitem[{{Axelrod} {et~al.}(2004){Axelrod}, {Connolly}, {Ivezic}, {Kantor},
  {Lupton}, {Plante}, {Stubbs}, \& {Wittman}}]{2004AAS...20510811A}
{Axelrod}, T., {Connolly}, A., {Ivezic}, Z., {et~al.} 2004, in Bulletin of the
  American Astronomical Society, Vol.~36, American Astronomical Society Meeting
  Abstracts, \#108.11

\bibitem[{{Barrett} \& {Bridgman}(1999)}]{pyfits}
{Barrett}, P.~E., \& {Bridgman}, W.~T. 1999, in Astronomical Society of the
  Pacific Conference Series, Vol. 172, Astronomical Data Analysis Software and
  Systems VIII, ed. D.~M. {Mehringer}, R.~L. {Plante}, \& D.~A. {Roberts}, 483

\bibitem[{{Boehm}(1981)}]{1981see..book.....B}
{Boehm}, B.~W. 1981, {Software Engineering Economics} (Prentice Hall)

\bibitem[{{Greenfield} \& {White}(2000)}]{pyraf}
{Greenfield}, P., \& {White}, R.~L. 2000, in Astronomical Society of the
  Pacific Conference Series, Vol. 216, Astronomical Data Analysis Software and
  Systems IX, ed. N.~{Manset}, C.~{Veillet}, \& D.~{Crabtree}, 59

\bibitem[{{Greenfield}~P.(2003)}]{numarray}
{Greenfield}~P., {Miller} J.~T., H. J. W.~R. 2003, PyCon Conference

\bibitem[{{Greisen}(2003)}]{aips}
{Greisen}, E.~W. 2003, Information Handling in Astronomy - Historical Vistas,
  285, 109

\bibitem[{Hunter(2007)}]{Hunter:2007}
Hunter, J.~D. 2007, Computing In Science \& Engineering, 9, 90

\bibitem[{{Jenness} {et~al.}(2016){Jenness}, {Bosch}, {Owen}, {Parejko},
  {Sick}, {Swinbank}, {de Val-Borro}, {Dubois-Felsmann}, {Lim}, {Lupton},
  {Schellart}, {Krughoff}, \& {Tollerud}}]{LSST&Astropy}
{Jenness}, T., {Bosch}, J., {Owen}, R., {et~al.} 2016, in Proc.~SPIE, Vol.
  99130G, Software and Cyberinfrastructure for Astronomy III

\bibitem[{{Joye} \& {Mandel}(2003)}]{ds9}
{Joye}, W.~A., \& {Mandel}, E. 2003, in Astronomical Society of the Pacific
  Conference Series, Vol. 295, Astronomical Data Analysis Software and Systems
  XII, ed. H.~E. {Payne}, R.~I. {Jedrzejewski}, \& R.~N. {Hook}, 489

\bibitem[{{Robitaille} \& {Bressert}(2012)}]{aplpy}
{Robitaille}, T., \& {Bressert}, E. 2012, {APLpy: Astronomical Plotting Library
  in Python}, Astrophysics Source Code Library, , , ascl:1208.017

\bibitem[{{Tody}(1986)}]{iraf}
{Tody}, D. 1986, in \procspie, Vol. 627, Instrumentation in astronomy VI, ed.
  D.~L. {Crawford}, 733

\bibitem[{{Tollerud}(2012)}]{astropysics}
{Tollerud}, E. 2012, {Astropysics: Astrophysics utilities for python},
  Astrophysics Source Code Library, , , ascl:1207.007

\bibitem[{{Turk} {et~al.}(2011){Turk}, {Smith}, {Oishi}, {Skory}, {Skillman},
  {Abel}, \& {Norman}}]{2011ApJS..192....9T}
{Turk}, M.~J., {Smith}, B.~D., {Oishi}, J.~S., {et~al.} 2011, ApJS, 192, 9

\bibitem[{{Van Der Walt} {et~al.}(2011){Van Der Walt}, {Colbert}, \&
  {Varoquaux}}]{numpy}
{Van Der Walt}, S., {Colbert}, S.~C., \& {Varoquaux}, G. 2011, {Computing in
  Science \& Engineering}, Vol 13, 22

\bibitem[{{Wallace} \& {Warren-Smith}(2000)}]{starlink}
{Wallace}, P.~T., \& {Warren-Smith}, R.~F. 2000, Information Handling in
  Astronomy, 250, 93

\bibitem[{{Weiner} {et~al.}(2009){Weiner}, {Blanton}, {Coil}, {Cooper},
  {Dav{\'e}}, {Hogg}, {Holden}, {Jonsson}, {Kassin}, {Lotz}, {Moustakas},
  {Newman}, {Prochaska}, {Teuben}, {Tremonti}, \&
  {Willmer}}]{2009astro2010P..61W}
{Weiner}, B., {Blanton}, M.~R., {Coil}, A.~L., {et~al.} 2009, in ArXiv
  Astrophysics e-prints, Vol. 2010, astro2010: The Astronomy and Astrophysics
  Decadal Survey

\end{thebibliography}

\bigskip
\bigskip
\bigskip
\section*{Affiliations}
\hypertarget{affil:1}{}
\noindent $^1$ {Center for Cosmology and AstroParticle Physics, The Ohio State University, 191 West Woodruff Avenue, Columbus, OH 43210, USA} \hypertarget{affil:2}{}\\
$^{2}$ {Department of Astronomy, The Ohio State University, 4055 McPherson Lab, 140 W 18th Ave, Columbus, OH 43210, USA} \hypertarget{affil:3}{}\\
$^{3}$ {Astrophysics Source Code Library} \hypertarget{affil:4}{}\\
$^{4}$ {Department of Physics, University of Warwick, Gibbet Hill Road, Coventry, CV4 7AL, UK} \hypertarget{affil:5}{}\\
$^{5}$ {European Southern Observatory, Alonso de Cordova 3107, Santiago, Chile} \hypertarget{affil:6}{}\\
$^{6}$ {Mullard Space Science Laboratory, Holmbury Hill Rd., Dorking, RH5 6NT Surrey, UK} \hypertarget{affil:7}{}\\
$^{7}$ {Astrophysics Research Centre, Queen's University Belfast, Belfast BT7 1NN, UK} \hypertarget{affil:8}{}\\
$^{8}$ {The Observatories of the Carnegie Institution for Science, 813 Santa Barbara St., Pasadena, CA 91101, USA} \hypertarget{affil:9}{}\\
$^{9}$ {Instituto de Fisica y Astronom\'{i}a, Facultad de Ciencias, Universidad de Valpara\'{i}so, Gran Bretan\~{a} 1111, Playa Ancha, Valpara\'{i}so, Chile} \hypertarget{affil:10}{}\\
$^{10}$ {Infrared Processing and Analysis Center, California Institute of Technology, Pasadena, CA 91125, USA} \hypertarget{affil:11}{}\\
$^{11}$ {Department of Astronomy \& Astrophysics, The Pennsylvania State University, University Park, PA 16802, USA} \hypertarget{affil:12}{}\\
$^{12}$ {Department of Physics and Astronomy, The University of Western Ontario, London, ON N6A 3K7, Canada } \hypertarget{affil:13}{}\\
$^{13}$ {Department of Physics, University of Oregon, Eugene, OR 97403, USA} \hypertarget{affil:14}{}\\
$^{14}$ {Instituto de Astrof\'{i}sica, Pontificia Universidad Cat\'{o}lica de Chile, Av.~Vicuna Mackenna 4860, 782-0436 Macul, Santiago, Chile} \hypertarget{affil:15}{}\\
$^{15}$ {School of Physics and Astronomy, Cardiff University, Cardiff, CF24 3AA, UK} \hypertarget{affil:16}{}\\
$^{16}$ {Department of Physics, Faculty of Engineering and Physical Sciences, University of Surrey, Guildford GU2 7XH} \hypertarget{affil:17}{}\\
$^{17}$ {Instituto de F\'{i}sica Te\'{o}rica (IFT) UAM/CSIC, Universidad Aut\'{o}noma de Madrid, Cantoblanco, E-28049 Madrid, Spain} \hypertarget{affil:18}{}\\
$^{18}$ {Centre for Astrophysics Research, Science and Technology Research Institute, University of Hertfordshire, Hatfield AL10 9AB, UK} \hypertarget{affil:19}{}\\
$^{19}$ {Centre for Astrophysics and Supercomputing, Swinburne University of Technology, Hawthorn, Victoria 3122, Australia} \hypertarget{affil:20}{}\\
$^{20}$ {Faculty of Science, Physics Department, University of Namibia, Windhoek, Namibia} \hypertarget{affil:21}{}\\
$^{21}$ {South African Astronomical Observatory (SAAO), Observatory Road Observatory Cape Town, WC 7925, South Africa} \hypertarget{affil:22}{}\\
$^{22}$ {Department of Physics \& Astronomy, Texas Christian University, Fort Worth, TX 76129, USA} \hypertarget{affil:23}{}\\
$^{23}$ {Universidade de S\~{a}o Paulo, Departamento de Astronomia do IAG/USP, Rua do Mat\~{a}o 1226, Cidade Universit\'{a}ria, 05508-900 S\~{a}o Paulo, SP, Brazil} \hypertarget{affil:24}{}\\
$^{24}$ {Columbia University, Department of Astronomy, 550 West 120th Street, Mail Code 5246, New York, NY, 10027, U.S.A.} \hypertarget{affil:25}{}\\
$^{25}$ {Department of Astronomy, University of Washington, Box 351580, Seattle, WA 98195, USA} \hypertarget{affil:26}{}\\
$^{26}$ {Center for Gravitation, Cosmology and Astrophysics, Department of Physics, University of Wisconsin Milwaukee, 3135 N Maryland Ave, Milwaukee, WI 53211, USA} \hypertarget{affil:27}{}\\
$^{27}$ {Instituto de Astronomia, Geof\'{i}sica e Ci\^{e}ncias Atmosf\'{e}ricas, Universidade de S\~{a}o Paulo, SP 05508-900, Brazil} \hypertarget{affil:28}{}\\
$^{28}$ {Institut Utinam, CNRS UMR 6213, Universit\'{e} de Franche-Comt\'{e}, OSU THETA Franche-Comt\'{e}-Bourgogne, Observatoire de Besan\c{c}on, BP 1615, 25010 Besan\c{c}on Cedex, France} \hypertarget{affil:29}{}\\
$^{29}$ {University of Geneva, Department of Astronomy, Chemin d'Ecogia 16, 1290, Versoix, Switzerland} \hypertarget{affil:30}{}\\
$^{30}$ {Institute of Cosmos Sciences, University of Barcelona, Barcelona, Spain} \hypertarget{affil:31}{}\\
$^{31}$ {Physics Department, University of Oxford, Denys Wilkinson Building, Keble Road, Oxford OX1 4RH, UK} \hypertarget{affil:32}{}\\
$^{32}$ {Max Planck Institute for Extraterrestrial Physics, Giessenbachstr.~1, 85748 Garching, Germany} \hypertarget{affil:33}{}\\
$^{33}$ {Joint ALMA Observatory, Alonso de C—rdova 3107, Vitacura, Santiago, Chile } \hypertarget{affil:34}{}\\
$^{34}$ {Space Telescope Science Institute, 3700 San Martin Dr., Baltimore, MD 21218, USA} \hypertarget{affil:35}{}\\
$^{35}$ {George Washington University, Staughton Hall, 707 22nd St NW, Washington, DC 20052, USA} \hypertarget{affil:36}{}\\
$^{36}$ {Astronomy Department, University of Massachusetts, Amherst, MA 01003, USA} \hypertarget{affil:37}{}\\
$^{37}$ {I.~Physikalisches Institut, University of Cologne, Z\"{u}lpicherstr.~77, 50937 Cologne, Germany} \hypertarget{affil:38}{}\\
$^{38}$ {Harvard-Smithsonian Center for Astrophysics, Cambridge MA 02138, USA} \hypertarget{affil:39}{}\\
$^{39}$ {Department of Physics, McGill University, 3600 rue University, Montr\'{e}al, QC H3A 2T8 Canada} \hypertarget{affil:40}{}\\
$^{40}$ {Department of Astronomy, University of Wisconsin, 2535 Sterling Hall, 475 N.~Charter Street, Madison, WI 53706, USA} \hypertarget{affil:41}{}\\
$^{41}$ {Department of Physics and Astronomy, York University, 4700 Keele St., Toronto, Ontario, Canada} \hypertarget{affil:42}{}\\
$^{42}$ {Institute of Cosmology and Gravitation, University of Portsmouth, Portsmouth PO1 3FX, UK} \hypertarget{affil:43}{}\\
$^{43}$ {School of Mathematics and Physics, University of Queensland, QLD 4072, Australia} \hypertarget{affil:44}{}\\
$^{44}$ {Princeton University, Peyton Hall, 4 Ivy Lane, Princeton, NJ 08540, USA} \hypertarget{affil:45}{}\\
$^{45}$ {University of Leiden, Sterrenwacht Leiden, Niels Bohrweg 2, NL-2333 CA Leiden, the Netherlands} \hypertarget{affil:46}{}\\
$^{46}$ {California Institute of Technology, 1200 East California Blvd, Pasadena, CA 91125, USA} \hypertarget{affil:47}{}\\
$^{47}$ {School of Earth and Space Exploration, Arizona State University, Tempe, AZ, 85287, USA} \hypertarget{affil:48}{}\\
$^{48}$ {Dark Cosmology Centre, Niels Bohr Institute, University of Copenhagen} \hypertarget{affil:49}{}\\
$^{49}$ {Instituto de Astrof\'{i}sica de Canarias (IAC), E-38205 La Laguna, Tenerife, Spain} \hypertarget{affil:50}{}\\
$^{50}$ {Leiden University, P.O.~Box 9513, 2300 RA Leiden, The Netherlands} \hypertarget{affil:51}{}\\
$^{51}$ {Argelander-Insitut for Astronomy, Auf dem Huegel 71, 53121 Bonn, Germany} \hypertarget{affil:52}{}\\
$^{52}$ {David Rittenhouse Laboratory, 209 South 33rd Street, Philadelphia, PA, 19104, USA} \hypertarget{affil:53}{}\\
$^{53}$ {Australian Astronomical Observatory, 105 Delhi Rd., North Ryde, NSW 2113 Australia} \hypertarget{affil:54}{}\\
$^{54}$ {Tartu Observatory, Observatooriumi 1, 61602 T\~{o}ravere, Estonia} \hypertarget{affil:55}{}\\
$^{55}$ {Department of Physics and Center for Gravitational Waves and Cosmology, West Virginia University, White Hall, Morgantown, WV 26506, USA} \hypertarget{affil:56}{}\\
$^{56}$ {Instituto de Astronomia, Universidad Nacional Aut\'{o}noma de M\'{e}xico, Apdo.~Postal 70-264, Cd.~Universitaria, CDMX 04510, M\'{e}xico} \hypertarget{affil:57}{}\\
$^{57}$ {School of Physics, University of Melbourne, Victoria 3010, Australia} \hypertarget{affil:58}{}\\
$^{58}$ {Department of Physics and Astronomy, University of Southampton, Southampton, SO17 1BJ, UK} \hypertarget{affil:59}{}\\
$^{59}$ {Physics Department, King's College, Strand, London, WC2R 2LS} \hypertarget{affil:60}{}\\
$^{60}$ {Centro de Investigaciones de Astronom\'{i}a (CIDA), Apartado Postal 264, M\'{e}rida 5101-A, Venezuela} \hypertarget{affil:61}{}\\
$^{61}$ {Instituto de F\'{i}sica e Qu\'{i}mica, Av.~BPS 1303 - Pinheirinho, Itajub\'{a} - MG - Brasil, CEP 37500-903} \hypertarget{affil:62}{}\\
$^{62}$ {Kapteyn Astronomical Institute, Rijksuniversiteit Groningen, Netherlands} \hypertarget{affil:63}{}\\
$^{63}$ {Yale-NUS College, 12 College Avenue West, Singapore 138614} \hypertarget{affil:64}{}\\
$^{64}$ {California State University, Chico, 400 W.~1st St.~Campus Box 0202, Chico, CA, USA} \hypertarget{affil:65}{}\\
$^{65}$ {Unidad de Astronom\'{i}a, Universidad de Antofagasta Avenida Angamos 601, Antofagasta 1270300, Chile} \hypertarget{affil:66}{}\\
$^{66}$ {Victorian Life Sciences Computation Initiative, Carlton, VIC, Australia} \hypertarget{affil:67}{}\\
$^{67}$ {Observatoire de Lyon, Universit«e Lyon 1, 9 Avenue Ch.~Andr\'{e}, Saint Genis Laval Cedex, France} \hypertarget{affil:68}{}\\
$^{68}$ {Physics Department, Lehigh University, 16 Memorial Drive East, Bethlehem, PA 18015, USA} \hypertarget{affil:69}{}\\
$^{69}$ {Department of Astrophysical Sciences, Princeton University, Princeton, NJ 08544, USA} \hypertarget{affil:70}{}\\
$^{70}$ {Macquarie University, Balaclava Road, Epping, NSW, Australia} \hypertarget{affil:71}{}\\
$^{71}$ {Department of Physical and Environmental Sciences, University of Toronto at Scarborough, Toronto, Ontario M1C 1A4, Canada} \hypertarget{affil:72}{}\\
$^{72}$ {Department of Astronomy, University of Michigan, 311 West Hall, 1085 South University Avenue, Ann Arbor, MI 48109, USA} \hypertarget{affil:73}{}\\
$^{73}$ {Institute of Theoretical Astrophysics, University of Oslo, PO 1029 Blindern, 0315 Oslo, Norway} \hypertarget{affil:74}{}\\
$^{74}$ {Department of Physics and Astronomy, Uppsala University, Box 516, 75120, Uppsala, Sweden} \hypertarget{affil:75}{}\\
$^{75}$ {Instituto de Ciencia de Materiales de Madrid (CSIC), E-28049, Madrid, Spain} \hypertarget{affil:76}{}\\
$^{76}$ {Florida Space Institute, University of Central Florida, 12354 Research Parkway, Orlando, FL, USA} \hypertarget{affil:77}{}\\
$^{77}$ {School of Physics \& Astronomy, University of St Andrews, North Haugh, St Andrews, KY16 9SS, UK} \hypertarget{affil:78}{}\\
$^{78}$ {Department of Physics and Astronomy, University College London, Gower Place, London WCE1 6BT, UK} \hypertarget{affil:79}{}\\
$^{79}$ {Institute of Astronomy, Universidad Nacional Aut\'{o}noma de M\'{e}xico, Ensenada, M\'{e}xico} \hypertarget{affil:80}{}\\
$^{80}$ {Max-Planck-Institut f\"{u}r Gravitationsphysik (Albert-Einstein-Institut), Am M\"{u}hlenberg 1, D-14476 Potsdam-Golm, Germany} \hypertarget{affil:81}{}\\
$^{81}$ {Yash Pal Center For Science And Technology, Anand Engineering College, Agra, India} \hypertarget{affil:82}{}\\
$^{82}$ {ASTRON, the Netherlands Institute for Radio Astronomy, Postbus 2, 7990 AA, Dwingeloo, The Netherlands} \hypertarget{affil:83}{}\\
$^{83}$ {Center for Astrophysics and Space Sciences (CASS), Department of Physics, University of California, San Diego, CA 92093, USA} \hypertarget{affil:84}{}\\
$^{84}$ {Institute of Astronomy, University of Cambridge, Madingley Road, Cambridge, CB3 0HA, UK} \hypertarget{affil:85}{}\\
$^{85}$ {Academia Sinica Institute of Astronomy \& Astrophysics (ASIAA), 11F of Astronomy-Mathematics Bldg, AS/NTU, No.~1, Roosevelt Rd.~Sec 4, Taipei 10617, Taiwan} \hypertarget{affil:86}{}\\
$^{86}$ {Anton Pannekoek Institute, University of Amsterdam, Postbus 94249, 1090 GE Amsterdam, The Netherlands} \hypertarget{affil:87}{}\\
$^{87}$ {Leibniz-Institut f\"{u}r Astrophysik Potsdam (AIP), An der Sternwarte 16, 14482 Potsdam, Germany} \hypertarget{affil:88}{}\\
$^{88}$ {Oak Ridge National Laboratory, Oak Ridge, TN 37831, USA} \hypertarget{affil:89}{}\\
$^{89}$ {University of Melbourne, School of Physics, Parkville 3010, Australia} \hypertarget{affil:90}{}\\
$^{90}$ {Center for Computational Sciences, University of Tsukuba, 1-1-1 Tennodai, Tsukuba, Ibaraki, 305-8577, Japan} \hypertarget{affil:91}{}\\
$^{91}$ {W.~M.~Keck Observatory, 65-1120 Mamalahoa Hwy.~Kamuela, HI, 96743, USA} \hypertarget{affil:92}{}\\
$^{92}$ {Department of Physics, University of California, Davis, CA, USA} \hypertarget{affil:93}{}\\
$^{93}$ {TAPIR, California Institute of Technology, Pasadena, CA, USA} \hypertarget{affil:94}{}\\
$^{94}$ {Department of Physics \& Astronomy, Johns Hopkins University, 3400 N.~Charles Street, Baltimore, MD, 21218, USA} \hypertarget{affil:95}{}\\
$^{95}$ {San Diego Supercomputer Center, UC San Diego, 9500 Gilman Drive, La Jolla, CA 92093, USA} \hypertarget{affil:96}{}\\
$^{96}$ {Department of Physics and Astronomy, University of Louisville, 102 Natural Science Building, Louisville KY 40292, USA} \hypertarget{affil:97}{}\\
$^{97}$ {Department for Astrophysics, University of Vienna, T\"{u}rkenschanzstr.~17, 1180, Vienna, Austria} \hypertarget{affil:98}{}\\
$^{98}$ {Department of Astronomy, Yale University, 52 Hillhouse Ave., New Haven, CT 06511, USA} \hypertarget{affil:99}{}\\
$^{99}$ {Institutionen f\"{o}r astronomi, Stockholms universitet, 106 91 Stockholm} \hypertarget{affil:100}{}\\
$^{100}$ {Astronomical Observatory, University of Warsaw, Al. Ujazdowskie 4, 00-478 Poland} \hypertarget{affil:101}{}\\
$^{101}$ {Dunlap Institute \& Department of Astronomy and Astrophysics, 50 St.~George Street, Toronto M5S 3H4, Canada} \hypertarget{affil:102}{}\\
$^{102}$ {Hamburg University, Gojenbergsweg 112, 21029 Hamburg, Germany} \hypertarget{affil:103}{}\\
$^{103}$ {Nicolaus Copernicus Astronomical Center, Warsaw, Bartycka 18, Poland} \hypertarget{affil:104}{}\\
$^{104}$ {Isaac Newton Group, Apartado 321, E-38700 S/C de La Palma, Spain} \\

\end{document}